\documentclass[aps,reprint,twocolumn,superscriptaddress,showpacs,pra,nofootinbib,floatfix]{revtex4-1}
\usepackage{amsmath}
 
 \usepackage{wasysym} 
 \usepackage{mathrsfs} 
 \usepackage{graphicx} 
 \usepackage{amsfonts} 
 \usepackage{amsbsy} 
 \usepackage{amscd} 
 \usepackage{amssymb}
\usepackage{bbm}  
 \usepackage{accents} 
 \usepackage{multirow} 
\usepackage{color}

\begin{document}
 
%\title{Thermally modified time travel of qubits}

\title{Distinguishing quantum states using time travelling qubits in a presence of thermal environments}

\author{Bartosz Dziewit}
\affiliation{Institute of Physics, University of Silesia in Katowice,  40-007 Katowice, Poland}
\author{Monika Richter}
\affiliation{Institute of Physics, University of Silesia in Katowice,  40-007 Katowice, Poland}
\author{Jerzy Dajka}
\affiliation{Institute of Physics, University of Silesia in Katowice,  40-007 Katowice, Poland}
\affiliation{Silesian Center for Education and Interdisciplinary Research, University of Silesia in Katowice, 41-500 Chorz\'{o}w, Poland}

\begin{abstract}
We consider quantum circuits with time travel designed for distinguishing specific non--orthogonal quantum states in two most popular models:  Deutsch's and postselected. We modify them by  a presence of weakly coupled thermal environment. Using the Davies approximation we study how the thermal noise affects an ability of the circuits to distinguish non--orthogonal quantum states. We show that for purely dephasing environment a 'paradoxial power' of such circuits remains preserved. We also present a physics--based argument for conditions of validity of the  maximum entropy rule introduced by David Deutsch for resolving the uniqueness ambiguity in a circuit with time travel.
\end{abstract}

\maketitle

\section{Introduction}

Impossibility of  distinguishing non--orthogonal quantum states is a bedrock  granting safety of quantum communication protocols~\cite{nielsen,scarani2,crypto}. This bedrock, however,  can be  eroded by  closed time--like curves (CTC) which existence (under certain assumptions) has already been predicted  long time ago~\cite{godel}. Potential time travelers  could utilize the 'paradoxial power' of such circuits to solve problems which are hard to solve  or even impossible to perform, cf. Ref. ~\cite{brun_exp} for recent a review. In particular they may be able to distinguish non--orthogonal quantum states~\cite{brun_disting,brun_fund}. 

There are at least three non--relativistic models, utilizing quantum circuit formalism,  of how the quantum computation is affected by the presence of CTCs. In other words,  there are at least three models of quantum time travel useful for quantum information. {\it (i)}  
 David Deutsch~\cite{deutsch} was the first who began to investigate properties of quantum systems in a presence of CTCs. He proposed an effective (nonrelativistic) description utilizing the quantum circuit formalism to describe quantum systems built of interacting the {\it chronology respecting} (CR) and {\it chronology violating} (CV) constituents. This proposal allowed to resolve at least some of the paradoxes caused by CTCs.  
Despite of experimental attempts of mimicking the Deutsch model~\cite{ralph,brun_exp} this proposal remains controversial~\cite{wal_fund,allen}. The second {\it (ii)},  utilizes a nowadays experimentally accessible teleportation protocol equipped with a post--selection~\cite{svet,seth_prl,seth_prd} and the third {\it (iii)}, most recent~\cite{allen}, uses transition probabilities.   For examples of other approaches one can consult e.g. Ref. ~\cite{elze_time} or ~\cite{vaidman_past}.

One can expect that the 'paradoxial' computational power of time travelers, originating from non--linearity of quantum models in the presence of CTCs, becomes weakened by omnipresent decoherence. In this paper we consider how the ability of distinguishing states of qubits is affected by thermal environment of time traveling qubit. We apply the Davies weak coupling approach~\cite{alicki} for a model of decoherence which is reviewed in Sec 2. of our paper.
We limit our attention quantum circuits distinguishing non--orthogonal qubit's states in the Deutsch model in Sec 3. of the paper and the post--selected teleportation in Sec. 4. In Sec. 5 we analyze the circuit for the unproven theorem~\cite{allen}, designed to exemplify a celebrated   paradox of information originating out of nowhere, and we present, utilizing general consideration of Ref.\cite{allen},  how the effect of thermal decoherence can serve as a physical justification of the Deutsch's maximum entropy rule introduced {\it ad hoc} in Ref.\cite{deutsch} in order to resolve the uniqueness ambiguity~\cite{allen} present  circuits with CTCs.       
    
%\section{Methods}  
\section{Davies decoherence}

Quantum decoherence  is  caused by the environment. Its influence  on the qubit $Q$  is modeled by the Hamiltonian  in the form:
\begin{eqnarray}\label{hamful}
H&=&H_Q+H_{env}+H_{int},
\end{eqnarray}
where $H_Q$ is the Hamiltonian of the qubit, $H_{env}$ models the environment and $H_{int}$ describes the qubit--environment interaction. 

For the qubit:
\begin{eqnarray}\label{hami}
H_Q&=&\frac{\omega}{2}(|1\rangle\langle 1|-|0\rangle\langle 0|), 
\end{eqnarray}
where $\omega$  is the energy splitting of the  qubit and $|0\rangle,|1\rangle$ span a Hilbert space of $Q$.  

We assume that the interaction between the qubit and its environment satisfies the Davies weak coupling conditions \cite{alicki} dedicated  for rigorous construction of the  qubit reduced dynamics calculated with respect to the environment. It is formulated  in terms of a completely positive (strictly Markovian) semigroup using  parameters of the  microscopic Hamiltonian of the full system~\cite{alicki}. As the Davies semigroups can be  rigorously and  consistently derived from  microscopic models of open systems they satisfy thermodynamic and statistical--mechanical properties of open quantum systems such as the detailed balance condition and the Gibbs canonical distribution in the stationary regime \cite{alicki}.
The Davies method  has been successfully  used in recent studies of various problems in quantum information and physics of open quantum systems including teleportation~\cite{kloda}, entanglement dynamics~\cite{lendi},  quantum discord~\cite{mymy,mymy2},   properties of geometric phases of qubits \cite{dav_faza},  thermodynamic properties of nano--systems \cite{dav_heat} and quantum games~\cite{dajka_game}.
In this paper we consider only   certain elements of Davies semi--groups:  the Davies {\it maps} $D=D(p,A,G,\omega,t)$ which acts  as follows~\cite{dav}:
%\begin{wid\section{Model of thermal environment}
%\begin{widetext}
%
\begin{equation}\label{dav}
\begin{array}{ll}
D\bigg[|1\rangle\langle 1|\bigg]= [1-(1-p)(1-e^{-At})]|1\rangle\langle 1|+ \\
+ (1-p)(1-e^{-At})|0\rangle\langle 0|, \\
D\bigg[|1\rangle\langle 0|\bigg]= e^{i\omega t -Gt}|1\rangle\langle 0|, \\
D\bigg[|0\rangle\langle 1|\bigg]= e^{-i\omega t -Gt}|0\rangle\langle 1|, \\
%\label{dav4}
D\bigg[|0\rangle\langle 0|\bigg]= p(1-e^{-At})|1\rangle\langle 1|+[1-(1-e^{-At})p]|0\rangle\langle 0| 
\end{array},
\end{equation}
%
%\end{widetext}
%
where $p\in[0,1/2]$ is related to the temperature $T$ of the environment via:
\begin{eqnarray}\label{p}
p&=&\exp(-\omega/2T)/[\exp(-\omega/2T)+\exp(\omega/2T)].
\end{eqnarray}
We set $k_B=1$.  
Let us notice that in long time limit the Davies map  transforms any qubit state $\rho$ into the equilibrium Gibbs state:
%%%%%%%%
\begin{eqnarray}\label{gibb}\lim_{t\rightarrow\infty}D(p,A,G,\omega,t)\rho=p|1\rangle\langle 1|+(1-p)|0\rangle\langle 0|.\end{eqnarray}
The case $T=0$ corresponds to the value $p=0$ and for $T\to \infty$ the parameter $p\to1/2$. 

The parameters $A = 1/\tau_R$ and $G = 1/\tau_D$ interpreted in terms of spin relaxation~\cite{T12}  are related to the
energy relaxation time $\tau_R$ and the dephasing time $\tau_D$ respectively~\cite{dav}.
There is a relation between $A$ and $G$ which guarantee that  the Davies  map  is a trace-preserving completely positive map. It is given by
the   inequalities~\cite{T12}
\begin{eqnarray}\label{warun}
G &\ge& A/2 \ge 0.%,  \quad \mbox{i.e.} \quad  2 \tau_1 \ge \tau_2
\end{eqnarray}
The limiting case  $A=0$ and $G\ne 0$ corresponds to (Markovian) pure dephasing  without dissipation of energy. Let us notice that the pure dephasing  despite its apparent simplicity can be effective applied to modeling of realistic systems c.f. Ref.\cite{defaz}, in which no energy dissipation occurs for a time scale significantly larger than other time scales in the system. 

%\section{Results and Discussion}
\section{Deutschian model of CTC}
The simplest Deutsch's circuit~\cite{deutsch} designed to mimic quantum dynamics in a presence of closed time--like curves (CTC) consists of a pair of qubits: the one is chronology respecting (CR) whereas the second violating the chronology (CV). This two qubits are coupled by the unitary $U$.  The CV qubit enters the circuit and interacts with the CR qubit. Then it {\it violates} the chronology and is identified with its past.  Formally the CV time evolution reads as follows:
\begin{equation}\label{d0}
\begin{array}{l@{}l}
\tau &{}=\Lambda(\tau),\\
\Lambda(\cdot) &{}=\mbox{Tr}_{CR}\{U(\rho_i\otimes\cdot)U^\dagger \},
\end{array}
\end{equation}
with the partial trace $\mbox{Tr}_{CR}$ calculated with respect to the CR qubit.

At the same time the state of the CR qubit which enters the circuit in a state $\rho_i$ changes into its final form $\rho_f$  given by:
\begin{eqnarray}\label{d2}
%\tau&=&\mbox{Tr}_{CR}\{U(\rho_i\otimes\tau)U^\dagger \}\\
\rho_f&=&\mbox{Tr}_{CV}\{U(\rho_i\otimes\tau)U^\dagger \}
\end{eqnarray}
with the partial trace  calculated with respect to the CV qubit.

Fundamentals of the Deutsch's consistence condition Eq.(\ref{d0}) is a subject of an important debate~\cite{wal_fund}. In this paper we do not intend to enter such philosophical topics and simply {\it assume} the ontic interpretation of quantum states both pure and, what is probably more unconventional, mixed. 

Instead, our aim is to investigate an effect of thermal noise affecting the CV qubit. We consider the Deutsch's consistence condition Eq.(\ref{d0}) modified by the presence of thermal environment in the Davies approximation discussed in the previous section. It is given by a composition $\circ$ of maps:
\begin{eqnarray}\label{d1}
\tau&=&[D\circ\Lambda](\tau),%\\
\end{eqnarray}

where $D=D(p,A,G,\omega,t)$ is the Davies map Eq.(\ref{dav}), with $t$ denoting a time period when the CV qubit interacts with thermal bath. Equation (\ref{d1}) has a natural interpretation: the CV qubit, before it returns to its past, interacts with thermal environment in the Markovian Davies approximation given by the map $D=D(p,A,G,\omega,t)$. Let us notice that the position of the $D$ in Eq.(\ref{d1}) rather than formal has a physical meaning reflecting our intention of making time travel 'noisy'. 

\begin{figure}[h!]
\begin{center}
\includegraphics[scale=0.4]{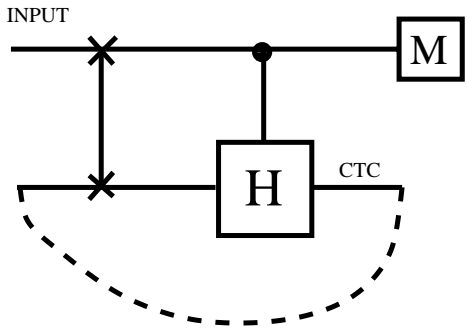}
\end{center}
\caption{Quantum circuit which can distinguish non--orthogonal states $|0\rangle$ and $|-\rangle$ using D--CTC (Deutschian). The dotted line denotes the qubit traveling backward in time. The circuit consists of the SWAP gate and the  controlled Hadamard $H$ gate. $M$ denotes a measurement of $\rho_f$ Eqs.(\ref{dd2}),(\ref{ddd2}). }
\label{fig1}
\end{figure}

Quantum circuits with CTCs can do tasks which are essentially inaccessible for the 'ordinary' (linear) quantum mechanics. One of the most spectacular examples of such a task is an ability of distinguishing non--orthogonal quantum states. This ability influences security of most quantum key distribution protocols~\cite{scarani2} with the celebrated archetype - the B92~\cite{B92}. There is a quantum circuit with the CTC~\cite{brun_disting} which can be utilized to distinguish non--orthogonal qubit states. It is presented in Fig.\ref{fig1}.
Formally its action is given by Eq.(\ref{d2}) and, in the presence of thermal environment, by Eq.(\ref{d1}) with the unitary $U$ given by

\begin{eqnarray}
U&=& |00\rangle\langle 00|+|01\rangle\langle 10|+|1+\rangle\langle 01|+|1-\rangle\langle 11|,
\end{eqnarray}
where $|\pm\rangle=[|0\rangle\pm|1\rangle]/\sqrt{2}$. It the noise--less case, when Eq.(\ref{d0}) instead of Eq.(\ref{d1}) is used, the circuit transforms the indistinguishable  states $|-\rangle$,$|0\rangle$ into  $|1\rangle$,$|0\rangle$ which are orthogonal and hence can be distinguished~\cite{brun_disting}. 
It is not surprising that the effect of thermal noise is to weaken this ability. 

In order to qualify an effect of noise we compare an output $\rho_f$ Eq.(\ref{d2}) of the noise--disturbed circuit with the noise--less output (which is $\xi_-=|1\rangle\langle 1|$ for $\rho_i=|-\rangle\langle -|$ and $\xi_0=|0\rangle\langle 0|$ for $\rho_i=|0\rangle\langle 0|$ respectively). 
We quantify an effect of noise
by the trace distance $Q(\rho_f,\xi)=\mbox{Tr}[\sqrt{(\rho_f-\xi)^2}]/2$~\cite{nielsen} which is known~\cite{nielsen} to indicate distinguishability the states $\rho_f,\xi$. For both inputs $\rho_i=|-\rangle\langle -|$ and $\rho_i=|0\rangle\langle 0|$ the corresponding states of the CV ($\tau$) and CR ($\rho_f$) qubits with the details of their calculation are given in the Appendix. 
%\begin{widetext}
For $\rho_i=|-\rangle\langle -|$ the trace distance $Q_-=Q(\rho_f,\xi_-)$ reads as follows
\begin{eqnarray}\label{Q1}
Q_-&=&  \frac{\sqrt {2}}{2}\,{\frac { \left( {{\rm e}^{A\,t}}-1 \right) {{\rm e}^{-G
\,t}}\sqrt {8\,{{\rm e}^{2\,G\,t}}+1} \left| -1+p \right| }{2\,{
{\rm e}^{A\,t}}-1}}.
\end{eqnarray}
For $\rho_i=|0\rangle\langle 0|$ the corresponding trace distance $Q_0=Q(\rho_f,\xi_0)$ is given by:
\begin{eqnarray}\label{Q2}
Q_0&=&  \frac{\sqrt {2}}{2}\,{\frac {p\, \left( {{\rm e}^{A\,t}}-1 \right) {{\rm e}^{
-G\,t}}\sqrt {8\,{{\rm e}^{2\,G\,t}}+1}}{2\,{{\rm e}^{A\,t}}-1}}.
\end{eqnarray}
%\end{widetext}

\begin{figure}[h!]
\begin{center}
\includegraphics[scale=0.4,angle=0]{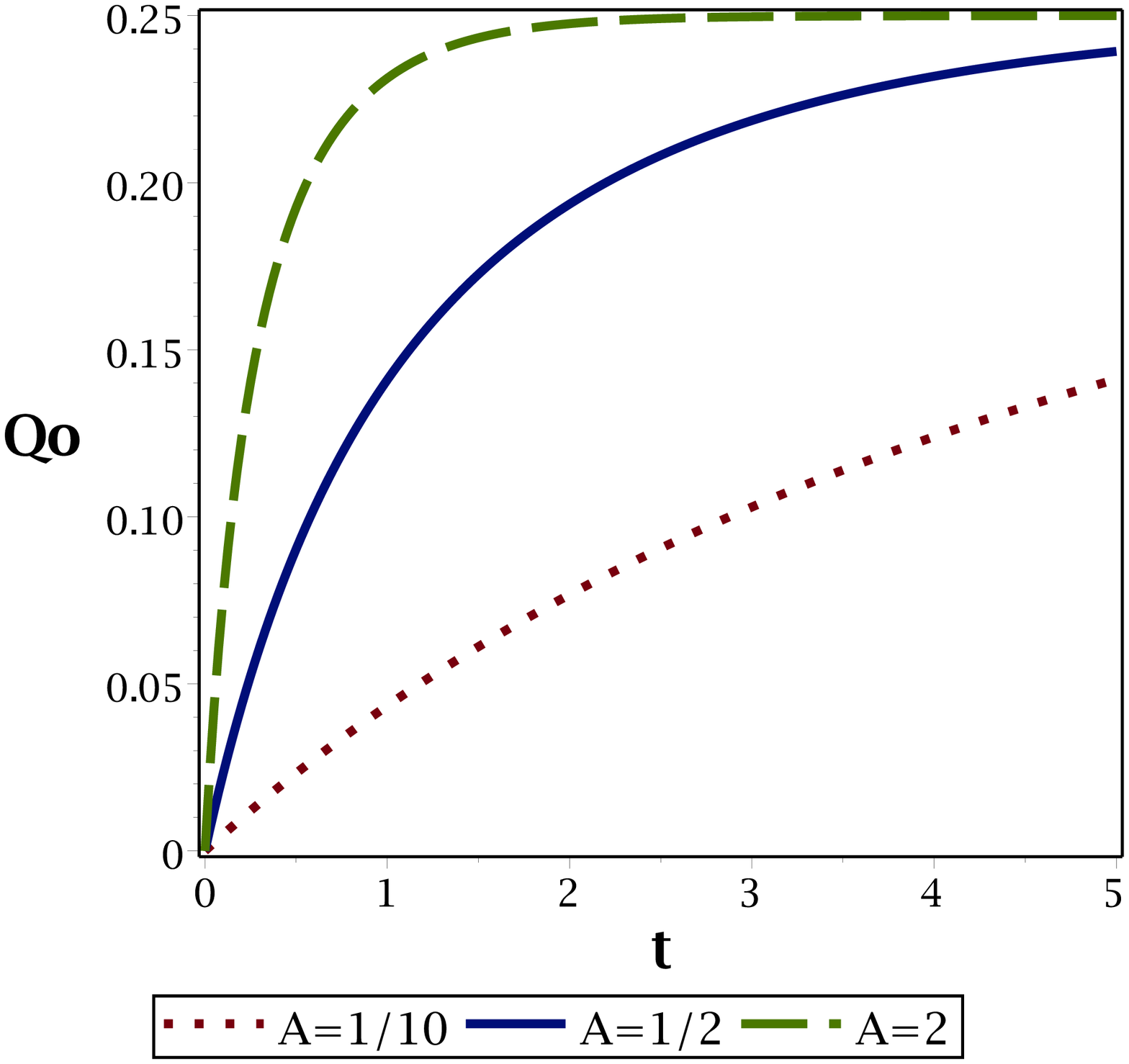}
\includegraphics[scale=0.4,angle=0]{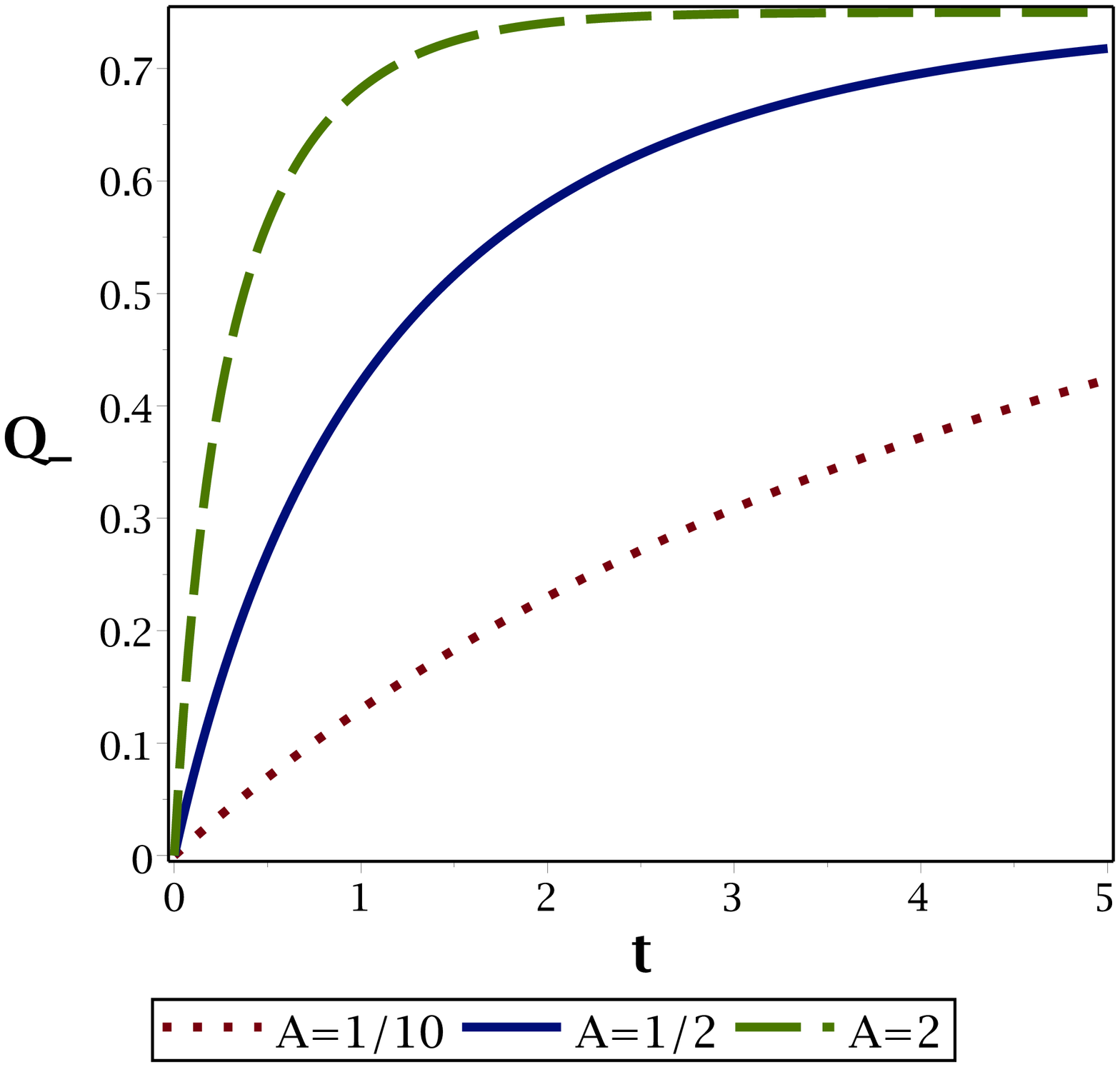}
\end{center}
\caption{Trace distance $Q$ calculated between the output of the circuit in Fig.(\ref{fig1}) for  the input $|0\rangle$ ($Q_0$ upper panel) and $|-\rangle$ ($Q_-$ lower panel) for different values of the parameter $A$ of the Davies map with $p=1/4$ and $G=1$}
\label{fig2}
\end{figure}

There are three parameters $p$, $A$ and $G$  describing thermal environment affecting the CV qubit via the Davies map. The first two affect qualitatively the value of $Q=Q_-,Q_0$. Increasing the last, $G$, has  only a quantitative impact and results in faster growth of both $Q_-$ and $Q_0$. It is not the case if one considers $A$.
The most important feature is that for {\it purely dephasing environments} the trace distance between the noisy and the noise--less output of the circuit in Fig.(\ref{fig1}) {\it vanishes} i.e. for $A=0$ both $Q_-=0$ and $Q_0=0$. In other words, in the case of pure dephasing the CTC--assisted distinguishing of non--orthogonal quantum states works as as good as in the  noise--less case.  Moreover, with decreasing $A$ the corresponding trace distance $Q_0,Q_-$ decreases as presented in Fig.(\ref{fig2}).    

\begin{figure}[h!]
\begin{center}
\includegraphics[scale=0.4,angle=0]{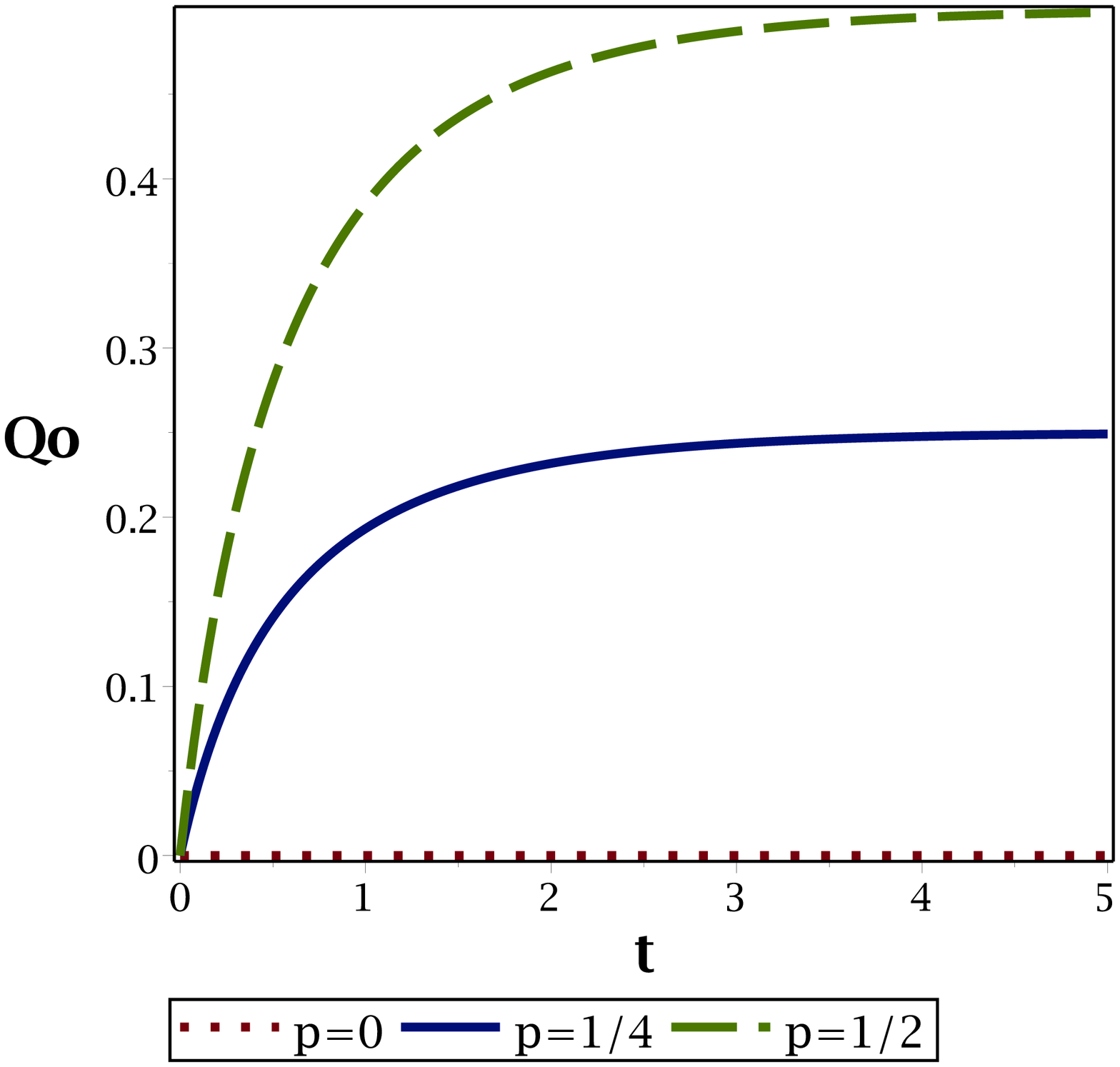}
\includegraphics[scale=0.4,angle=0]{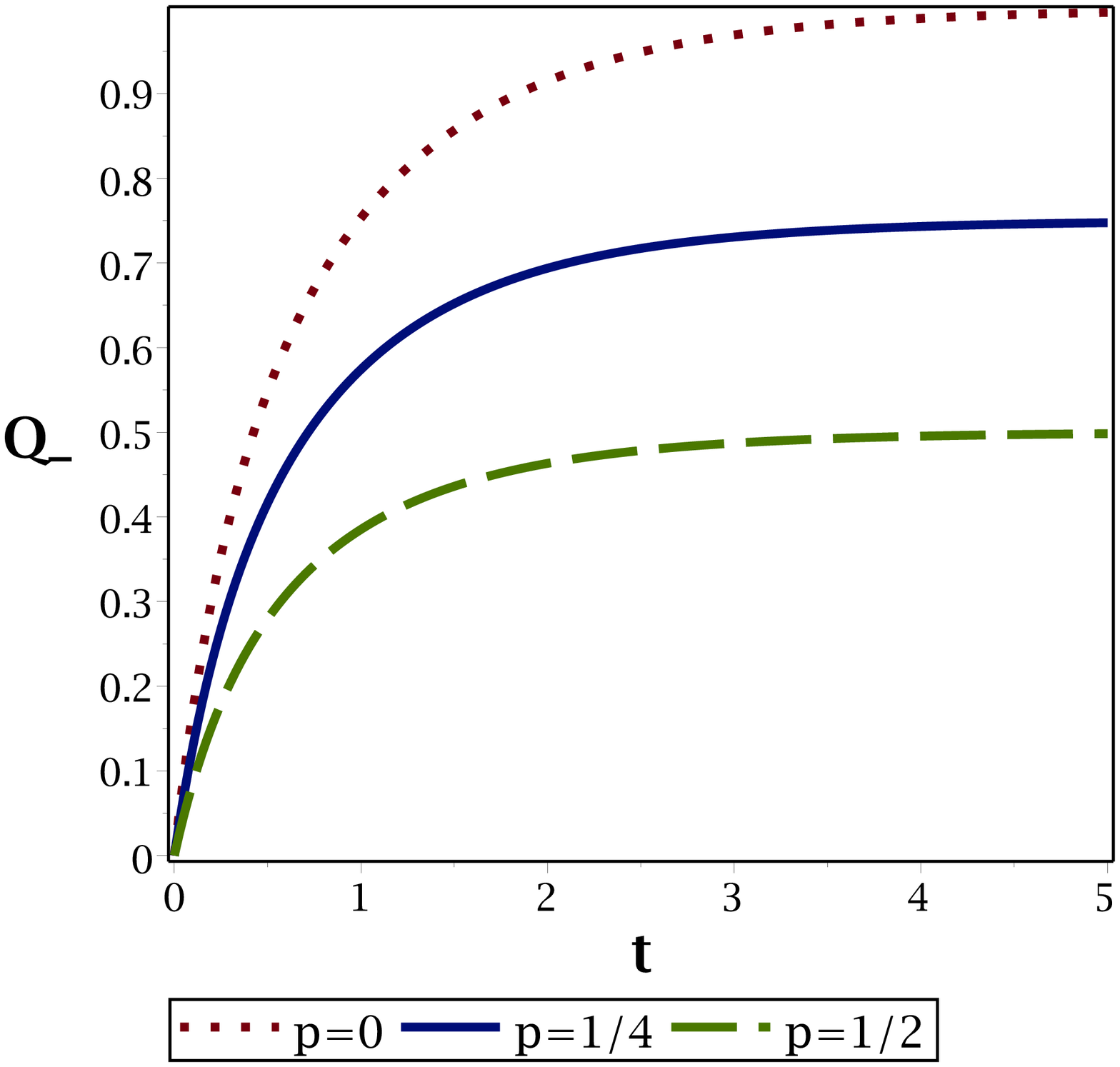}
\end{center}
\caption{Trace distance $Q$ calculated between the outputs of the circuit in Fig.(\ref{fig1}) for  the input $|0\rangle$ ($Q_0$ upper panel) and $|-\rangle$ ($Q_-$ lower panel) for different values of the parameter $p$ of the Davies map and $A=G=1$}
\label{fig3}
\end{figure}

Let us also notice that in the low temperature limit $p=0$ the trace distance $Q_0=0$ and that for larger values of $p$ the trace distance $Q_-$  {\it grows slower} than $Q_0$. As one infers from Fig.(\ref{fig3}) for fixed time  instant $t$ and ordered values $p_1<p_2$ the corresponding time derivatives $\partial Q_0/\partial t|_{t,p_1}<\partial Q_0/\partial t|_{t,p_2}$ whereas $\partial Q_-/\partial t|_{t,p_1}<\partial Q_-/\partial t|_{t,p_2}$. 
This seemingly counter--intuitive property results from the particular and distinguished role played by the pure dephasing limit and related symmetry~\cite{alidef}.  
\begin{figure}[h!]
\begin{center}
\includegraphics[scale=0.4,angle=0]{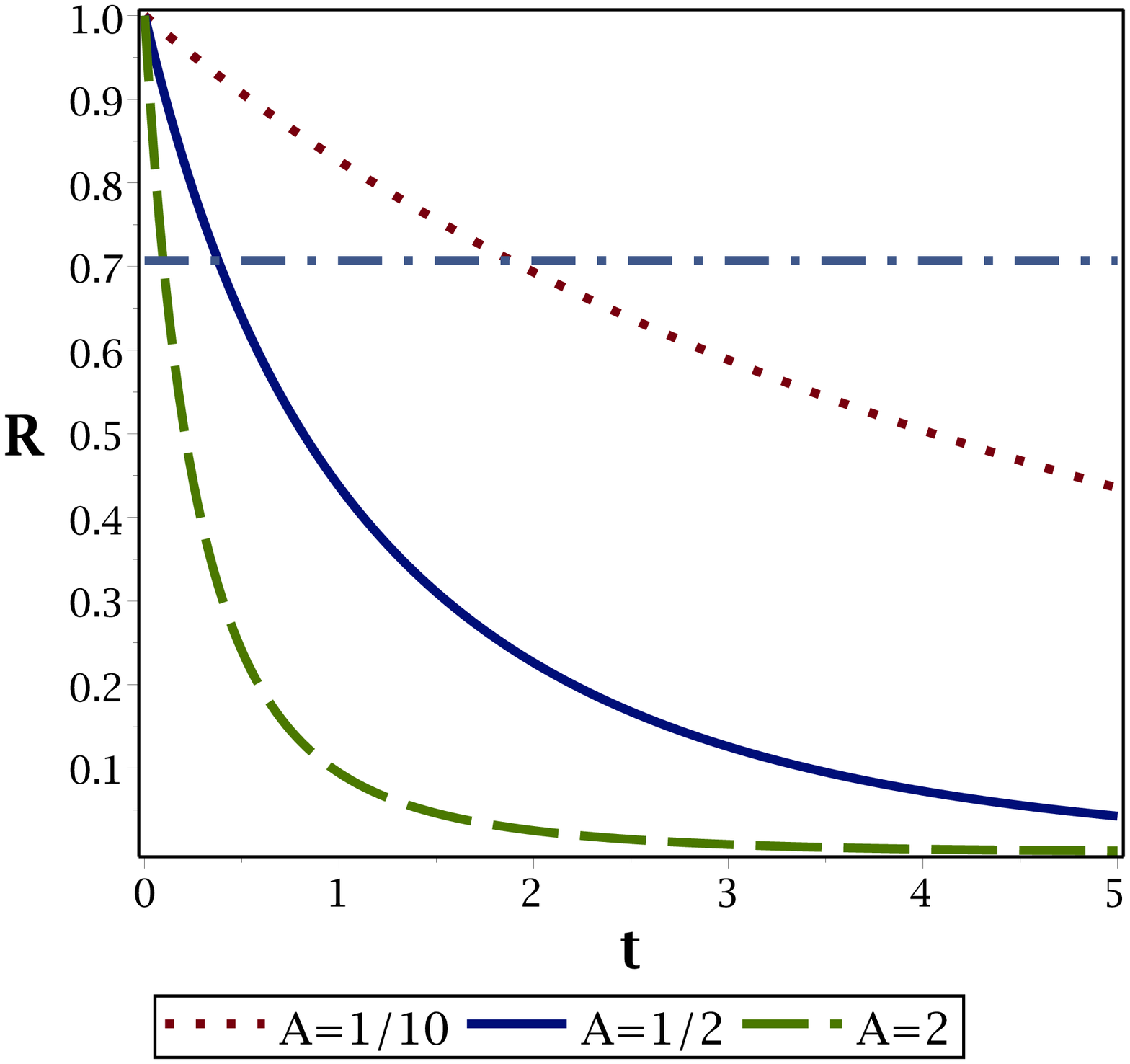}
\includegraphics[scale=0.4,angle=0]{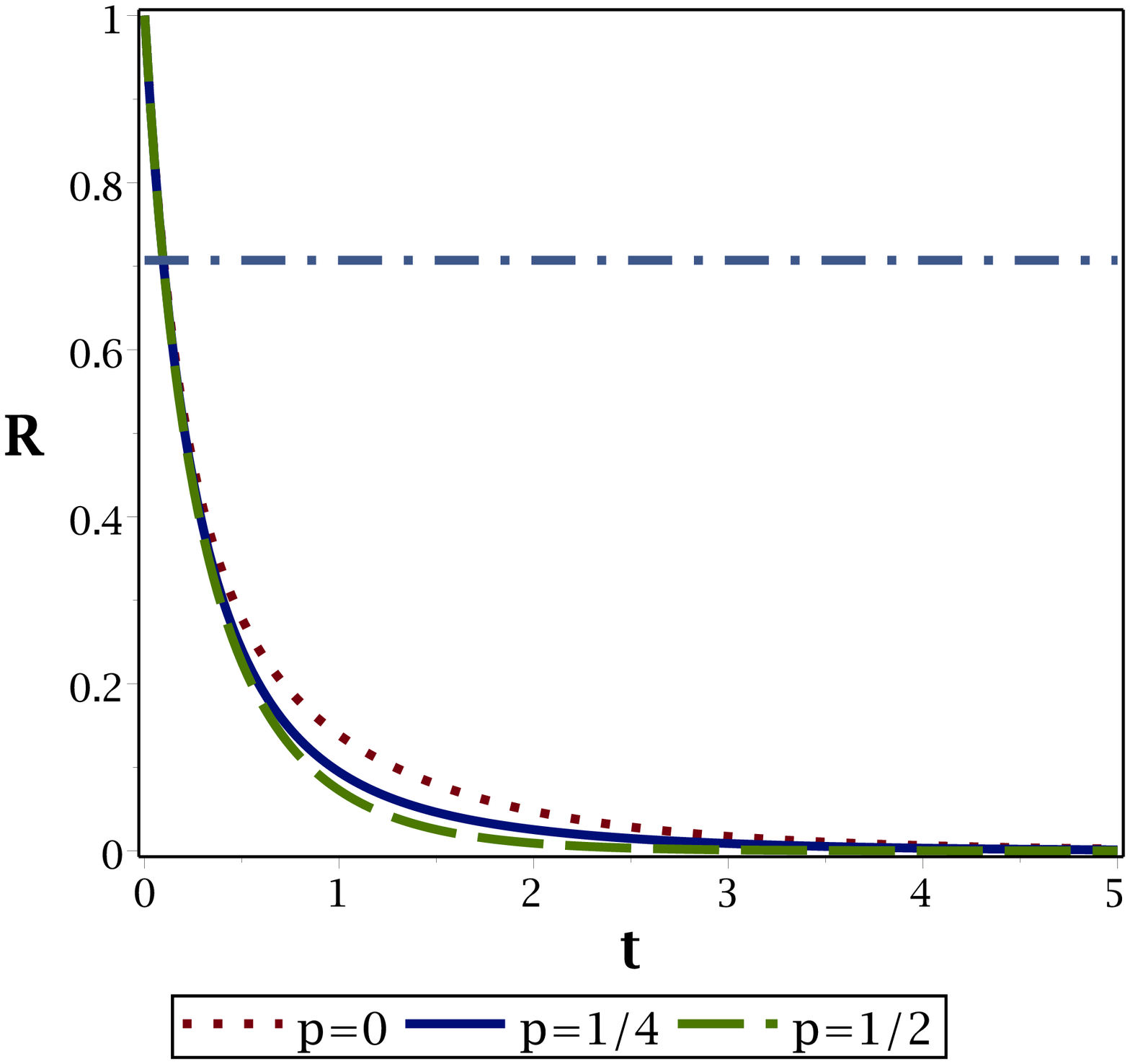}
\end{center}
\caption{
Trace distance $R$ Eq.(\ref{QQQ}) calculated between the output of the circuit in Fig.(\ref{fig1}) for  the thermally modified states Eq.(\ref{d2}) for the inputs $|0\rangle$ and $|-\rangle$  for different values of the parameter $A$, $p=1/4$ and $G=1$ (upper panel) and  for different values of the parameter $p$ and $A=2G=2$ (lower panel) of the Davies map. The horizontal line on both panels indicate $R(\rho_{i1},\rho_{i2})=\sqrt{2}/2$.}
\label{fig34}
\end{figure}

A natural quantifier of an effect of thermal environment on the 'paradoxial' power of distinguishing non--orthogonal states is a difference between the trace distance of two inputs $R(\rho_{i1},\rho_{i2})$ and the corresponding outputs $R(\rho_{f1},\rho_{f2})$ for $\rho_{f1},\rho_{f2}$ calculated via Eq.(\ref{d2}). As the circuit in Fig.(\ref{fig1}) is dedicated to distinguish two very particular states $|0\rangle,|-\rangle$, cf. Ref.~\cite{brun_disting}, with $R(\rho_{i1},\rho_{i2})=\sqrt{2}/2$, the figure of merit is the quantity $R(\rho_{f1},\rho_{f2})$ which reads as follows: 
\begin{eqnarray}\label{QQQ}
R&=& \frac{e^{-tG}}{4e^{At}-1}[ (1-4p+4p^2)[2e^{2At}-4e^{At}+2]%\nonumber \\
 + 4e^{2tG}]\nonumber \\
\end{eqnarray}
For $R=1$ the output states $\rho_{f1},\rho_{f2}$ are distinguishable. The smaller value of $R$ is the more ineffective  the circuit in Fig.(\ref{fig1}) is. Let us notice that for $R<\sqrt{2}/2$ distinguishability of the output states becomes, due to Davies decoherence, even worse than initially. The threshold condition  $R(\rho_{f1},\rho_{f2})=\sqrt{2}/2$, indicated by the horizontal line in Fig.(\ref{fig34}), depends not only on time instant $t$ but also on parameters of the system. Decreasing $A$ (for given $G$ and $p$) allows to keep the circuit useful despite longer exposition on decoherence.  
Again, for $A=0$ the $R=1$ i.e. the output states in a presence of a purely dephasing environment are as good distinguishable as they where in a absence of decoherence as presented in Fig.(\ref{fig34}). It is natural to attempt to generalize this result beyond a limited class of input states which is the circuit in Fig.(\ref{fig1}) designed for. Although we cannot present a formal proof, we conjecture, upon numerical experiments performed on randomly chosen pairs of non--orthogonal initial states, that a thermal environment never enhance state distinguishability which is 'best' in the pure dephasing limit $A=0$.  It is known that non--completely positive maps describing e.g. time--evolution of quantum systems initially entangled with their environment are not contractive~\cite{distance,laine}. As the Davies map Eq.(\ref{dav}) is, under the condition Eq.(\ref{warun}), contractive, one expects that any enhancement of distinguishability is solely due to peculiar character of the Deutsch map Eq.(\ref{d2}) and Eq.(\ref{d0}) originating from its non--linearity. 

\section{Post--selected CTC and thermal noise}
\begin{figure}[h!]
\begin{center}
\includegraphics[scale=0.4]{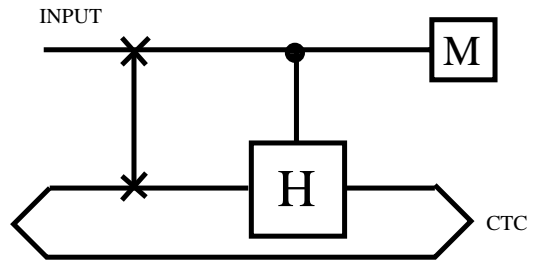}
\end{center}
\caption{Quantum circuit which can distinguish non--orthogonal states $|0\rangle$ and $|-\rangle$ using P--CTC (post--selected). The CV qubits are initially prepared in a maximally entangled state $|\Phi\rangle=[|00\rangle+|11\rangle]/\sqrt{2}$ which is then sent in the past by postselection of a projective $|\Phi\rangle\langle\Phi|$ measurement output. The other elements of the circuit are the same as in Fig.(\ref{fig1}).}
\label{fig4}
\end{figure}

The Deutsch's model~\cite{deutsch} of time travel  operates essentially beyond standard quantum mechanics. However, there is the second most popular circuit--based  model of quantum dynamics in a presence of CTCs in which one {\it mimics} the CV motion by a post--selected teleportation~\cite{svet,seth_prl,seth_prd,brun_fund}. Contrary to various difficulties arising in attempts of implementing Deutsch model~\cite{ralph,brun_exp} there are no fundamental experimental obstructions to post--select a desired outcome of teleportation procedure. However, let us notice that this apparent simplification occurs at cost of {\it deterministic post--selection} introduced {\it ad hoc} leading  the well defined quantum teleportation protocol out of quantum mechanics {\it per se}. In Fig. (\ref{fig4}) we present a well known circuit designed to transform (in an absence of noise) non--orthogonal states $|1\rangle$ and $|-\rangle$ into a  pair of orthogonal, and hence distinguishable, states $|1\rangle$ and $|0\rangle$, respectively~\cite{brun_fund}. 

The only but crucial difference between circuits in Fig(\ref{fig4}) and Fig.(\ref{fig1}) is in a way how an evolution of the CV qubit is modeled. 
Mimicking CTC with a post--selected teleportation utilizes a maximally entangled state as a resource~\cite{brun_fund} which, however, can be imperfect due to a presence of thermal noise.   Here we consider a state of two qubits and we assume that only one of parties in this resource is affected by thermal environment. Let us notice that such a setting is {\it physically} different to that which we adopt in previous studies of the Deutsch model where the time travel itself was assumed to be 'noisy'. 
Postselected CTC with thermal noise  affecting the maximally entangled Bell state $|\Phi\rangle=[|00\rangle+|11\rangle]/\sqrt{2}$ of the CV qubits, cf. Fig.(\ref{fig4}),
is given by
\begin{eqnarray}\label{p1}
\rho_f&=& \mbox{Tr}_{CB}\{|\Phi\rangle\langle\Phi|_{CB}U[\rho_i\otimes\chi_{CB}]U^\dagger \},
%\chi_{AB}&=&  D[|\Phi\rangle\langle\Phi|_{AB}]
\end{eqnarray}
where
\begin{eqnarray}\label{p2}
%\varrho_f&=& \mbox{Tr}_{AB}\{|\Phi\rangle\langle\Phi|_{AB}U[\rho_i\otimes\chi_{AB}]U^\dagger \}
\chi_{CB}&=&  [D\otimes I]|\Phi\rangle\langle\Phi|_{CB},
\end{eqnarray}
is the noisy Bell state obtained tensor product $D\otimes I$ of the Davies and an identity map.  It is assumed that  only the CV qubit in $\chi_{CB}$ labeled by $C$ is coupled to the thermal Davies environment. Let us notice formal analogy of this scenario with a recently studied thermally modified teleportation protocol~\cite{kloda} or entanglement swapping~\cite{mymy2}. 
In particular 
\begin{eqnarray}\label{pp2}
%\varrho_f&=& \mbox{Tr}_{AB}\{|\Phi\rangle\langle\Phi|_{AB}U[\rho_i\otimes\chi_{AB}]U^\dagger \}
\chi_{CB}&=& a_0|00\rangle\langle 00|_{CB}+b_0|10\rangle\langle 10|_{CB}\nonumber \\
&+& c^*|00\rangle\langle 11|_{CB}+c|11\rangle\langle 00|_{CB} \nonumber \\
&+&a_1|01\rangle\langle 01|_{CB}+b_1|11\rangle\langle 11|_{CB}
\end{eqnarray}
where
\begin{eqnarray}\label{dfun}
2b_1&=&1-(1-p)(1-e^{-At})\\
2a_1&=&(1-p)(1-e^{-At})\\
2c&=&e^{-i\omega t -Gt}\\
2b_0&=&p(1-e^{-At})\\
2a_0&=&1-(1-e^{-At})p
\end{eqnarray}
For $\rho_i=|\psi\rangle\langle \psi|$ pure, the action of the circuit in Fig.({\ref{fig4}) is, in the presence of Davies environment Eq.(\ref{p1}) is given by the following transformation:
\begin{eqnarray}\label{p1}
\rho_f&=& \langle\Phi|_{CB}U[|\psi\rangle\langle \psi|\otimes\chi_{CB}]U^\dagger|\Phi\rangle_{CB}\nonumber\\
&=& \frac{a_0}{2}L_{I}|\psi\rangle\langle \psi|L_I^\dagger+\frac{b_0}{2}L_{II}|\psi\rangle\langle \psi|L_{II}^\dagger\nonumber \\
&+& \frac{c^*}{2}L_{III}|\psi\rangle\langle \psi|L_{IV}^\dagger+ \frac{c}{2}L_{IV}|\psi\rangle\langle \psi|L_{III}^\dagger\nonumber \\
&+&\frac{a_1}{2}L_{V}|\psi\rangle\langle \psi|L_V^\dagger+\frac{b_1}{2}L_{VI}|\psi\rangle\langle \psi|L_{VI}^\dagger
%\chi_{AB}&=&  D[|\Phi\rangle\langle\Phi|_{AB}]
\end{eqnarray}
where (notice that $U=U_{SYS,C}$),
\begin{eqnarray}\label{ls}
L_I &=& \langle \Phi|_{CB} U|00\rangle_{CB}=\frac{1}{\sqrt{2}} |0\rangle \langle 0|\nonumber \\
L_{II} &=& \langle \Phi|_{CB} U|10\rangle_{CB} = \frac{1}{2}\left(|1\rangle\langle 0|+ |1\rangle\langle 1|\right)\nonumber \\
L_{III} &=& \langle \Phi|_{CB} U|00\rangle_{CB}=L_I\nonumber \\
L_{IV} &=& \langle \Phi|_{CB} U|11\rangle_{CB} = \frac{1}{2}\left(|1\rangle\langle 0|- |1\rangle\langle 1|\right)\nonumber \\
L_V &=& \langle \Phi|_{CB} U|01\rangle_{CB}=\frac{1}{\sqrt{2}} |0\rangle \langle 1|\nonumber \\
L_{VI} &=& \langle \Phi|_{CB} U|11\rangle_{CB} =L_{IV}
\end{eqnarray}

In a general case the transformation in Eq.(\ref{p1}) transforms non--orthogonal states into the states which remain non--orthogonal i.e. thermal Markovian noise divests the circuit in Fig.(\ref{fig4}) of its 'paradoxial' power (below we skip normalization constants):
%\begin{widetext}
%\begin{eqnarray}|1\rangle\langle 1|&\longrightarrow& \left[ \begin {array}{cc} \left( 1-p \right)  \left( 1-{{\rm e}^{-At}} \right) &0\\ \noalign{\medskip}0&p \left( 1-{{\rm e}^{-At}} \right)/2 +1/2-\left( 1-p \right)  \left( 1-{{\rm e}^{-At}}/2 \right) \end {array} \right] \\|+\rangle\langle +|&\longrightarrow& \left[ \begin {array}{cc} 1/2-p \left( 1-{{\rm e}^{-At}}\right)/2 +\left( 1-p \right)  \left( 1-{{\rm e}^{-At}} \right)/2 &0\\ \noalign{\medskip}0&p \left( 1-{{\rm e}^{-At}} \right) \end {array} \right] \end{eqnarray}
%\end{widetext}
%\begin{widetext}
\begin{eqnarray}
|1\rangle\langle 1|&\rightarrow&  \frac{1}{2}\left( 1-p \right)  \left( 1-{{\rm e}
^{-At}} \right) |0\rangle\langle 0| \nonumber \\&+&\frac{1}{2}\left[1+(2p-1) \left( 1-{{\rm e}^{-At
}} \right)\right]|1\rangle\langle 1|\nonumber \\
 \\
|+\rangle\langle +|&\rightarrow& \frac{1}{2}\left[ 1+(1-2p) \left( 1-{{\rm e}^{-At}}
 \right) \right] |0\rangle\langle 0| \nonumber \\ &+&\frac{1}{2}p \left( 1-{{\rm e}^{-At}} \right)|1\rangle\langle 1| 
\end{eqnarray}
%\end{widetext}
However, if  an energy exchange between the CV qubit and the environment is negligible, i.e. the circuit operates in the  pure dephasing regime $A=0$, the situation changes. Non--orthogonal input states are transformed into an output states which  {\it are orthogonal} and hence can be distinguished:    
\begin{eqnarray}\label{post}
|1\rangle\langle 1| &\longrightarrow&  |1\rangle\langle 1|\\
|+\rangle\langle +| &\longrightarrow&  |0\rangle\langle 0|
\end{eqnarray} 
The reason of that becomes clear if one notices that {\it both} in the original noise--less case~\cite{brun_fund} (i.e. when $D=I$) and in the   pure decoherence limit $A=0$
\begin{eqnarray}
a_1&=&b_0=0\\
a_0 &=& b_1 =1
\end{eqnarray}
 the transformation of non--orthogonal into the orthogonal states occurs since it follows that either $L_i|1\rangle\rightarrow |1\rangle$ and $L_i|+\rangle \rightarrow |0\rangle$ or $L_i|1\rangle =L_i|+\rangle=0$ for $i=I,\ldots,VI$.

In the above equations we used Eqs (\ref{p1}) and (\ref{p2}) but skip (non-vanishing) normalization constants which does not affect orthogonality of states. From Eq.(\ref{post}) one infers that also in the case of postselected teleportation model pure dephasing plays a distinguished role exactly as it was in the Deutsch's model.

\section{Uniqueness ambiguity}
According to the Schauder's fixed point theorem, there is a solution $\tau$ of the Deutsch's condition Eq.(\ref{d0}). However, such a solution may not be unique resulting in the {\it uniqueness ambiguity}~\cite{deutsch,allen}. Using the Deutsch model of quantum  time travel one faces with a problem which state $\tau$ (among many possibilities) is the 'proper' one. The original proposal of David Deutsch~\cite{deutsch} is the {\it maximum entropy rule} which states that the physical $\tau$ is the one which contains minimum information. This condition  introduced {\it ad hoc}~\cite{allen}  is not universal and  can be replaced by other proposals~\cite{politzer,dejonghe}.  As an example of a Deutsch's circuit with uniqueness ambiguity can serve a circuit designed for the unproven theorem paradox. It is an example of a knowledge--generating circuit: a mathematician $M$, equipped with a knowledge about her/his modern mathematics read from a book $B$, becomes a time traveler $T$ and travels back in time in order to write the book $B$.  A simplest example of a circuit playing such a role is  presented in Fig.(\ref{fig5}), cf. Ref.~\cite{allen}.
\begin{figure}[h!]
\begin{center}
\includegraphics[scale=0.4]{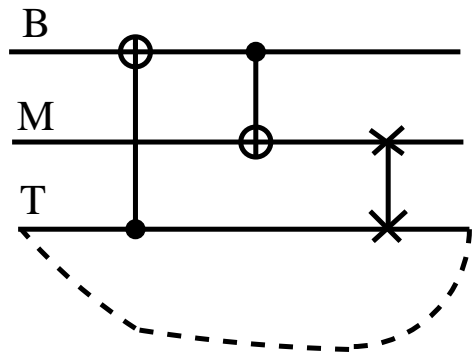}
\end{center}
\caption{Quantum circuit for the unproven theorem~\cite{allen}. $B$ is the book,  $M$ -- the mathematician and $T$ the time traveler using D--CTC and the action of the circuit is given in Eq.(\ref{book}).}
\label{fig5}
\end{figure}
Such a circuit describes interaction of three qubits:  $B$, $M$ which are CR and   the last one  $T$ which  violates chronology. The interaction is given by a unitary             
\begin{eqnarray}\label{book}
U&=& SWAP_{MT}CNOT_{BM}CNOT_{TB},
\end{eqnarray}
and an input of the circuit is $|0\rangle_B|0\rangle_M$. 

The Deutsch's consistency condition Eq.(\ref{d0}) for this circuit is solved by a family of states
\begin{eqnarray}\label{tau_b0}
\tau_\alpha &=& \alpha |0\rangle\langle 0| +(1-\alpha)|1\rangle\langle 1|,
\end{eqnarray}
where $\alpha\in[0,1]$ and hence is ambiguous. 
In Ref.\cite{allen} it is shown that an effect of depolarization can resolve this ambiguity. 

Here we consider probably the most natural and omnipresent source of noise. We assume that the time travel is disturbed by a thermal environment. In such a case the state of the time traveler is a solution of Eq.(\ref{d1}) i.e. the time travel of $T$ is affected by thermal Davies noise. This solution is unique and is given by the Gibbs state:    
\begin{eqnarray}
\label{tau_b}
% \tau = -{\frac {-{{\rm e}^{-At}}-p+p{{\rm e}^{-At}}}{1+{{\rm e}^{-At}}}}|0\rangle\langle 0| +\\+{\frac {1-p+p{{\rm e}^{-At}}}{1+{{\rm e}^{-At}}}} |1\rangle\langle 1|.
 \tau &=& p|1\rangle\langle 1|+(1-p) |0\rangle\langle 0|.
\end{eqnarray}
Let us notice that in the zero--temperature limit $p=0$ 
%or the unitary limit of the Davies map ($A\rightarrow 0$ {\it followed by} $G\rightarrow 0$ due to the constraint in Eq.(\ref{warun})) 
one obtains $\tau\rightarrow\tau_{0}$. In the  $p=1/2$ limit one arrives at the state which {\it maximizes} entropy. Let us also notice that, for the model of Davies decoherence considered here, the Deutsch's rule  holds only approximately (in
the regime of high temperature) and that in general the unique
solution is not always maximally mixed.  

The solution to the uniqueness ambiguity discussed here is essentially the
same as in Ref. \cite{allen}, but the source of noise that resolves the ambiguity is physically rather than formally motivated.   In other words, a very natural condition that the CV qubit is weakly disturbed by its thermal environment can serve as {\it physics--based} justification for the choice of the solution $\tau$ of the Deutsch consistency condition Eq.(\ref{d0}), instead of  otherwise {\it ad hoc}, maximum entropy rule introduced by David Deutsch in Ref.\cite{deutsch}.
\section{Summary}
If time travels were possible, the world would be essentially different. 
Quantum cryptography~\cite{crypto} and in particular quantum key distribution~\cite{scarani2} essentially changed basic objectives of communication which, comparing to a pre--quantum age, became much safer. However, most of the quantum no--go theorems -- {\it sine qua non} conditions for security of quantum protocols\cite{scarani1,scarani2} -- originate from {\it linearity} of quantum mechanics\cite{nielsen}. An existence of closed time--like curves can change (almost) everything. There are quantum circuits which in a presence of CTCs can break security of quantum protocols. In this work we analyzed only one of them: the circuit designed to distinguish non--orthogonal qubit's states and to  break e.g. the B92 quantum crypto--protocol. Our aim was to check if and how such a 'paradoxial power' becomes reduced by the omnipresent decoherence caused by thermal environment affecting time--traveling qubits. We consider only two among many approaches to CTCs: the one proposed by David Deutsch~\cite{deutsch} and the second based on the post--selected teleportation protocol~\cite{svet,seth_prl,seth_prd}.  Our intention was to investigate possibly wide class of open systems modeled in way which is both tractable and rigorous. That is why we assumed the  general type of coupling to the environment: the Davies weak coupling approach~\cite{alicki}. Using Davies approach one can describe the broadest class of open quantum systems with finite--dimensional space of states with {\it the only restriction} imposed: the coupling to environment must be weak.   We showed  for both  Deutsch's and post--selected model a distinctive role played by {\it pure decoherence} when, despite of a presence of environment and resulting information loss, circuits with CTCs do not lose their 'paradoxial power' of distinguishing non--orthogonal quantum states. This result can serve as a potentially useful guideline for experimentalists who attempt to mimic circuits with CTCs in order to implement 'linearity--free quantum computations'. Physically pure decoherence describe open quantum systems operating at time scales which are short comparing with a time scale of a system--environment energy exchange~\cite{defaz}. 

In addition to practical there is also a fundamental aspect of decoherence which needs to be taken into account in all the applications of quantum phenomena~\cite{schloss}. In the last section of our paper, inspired by Ref. \cite{allen}, we investigated the circuit for an unproven theorem to show that thermal decoherence, present in any real system, can help to resolve the uniqueness ambiguity originating from non--uniqueness of a solution of the Deutch's consistence condition Eq.(\ref{d0}). We showed that in a particular case considered in our work thermal noise not only allows to select the 'proper' state of chronology violating qubit, which is not necessarily maximally mixed,  but also justifies the Deutsch's maximum entropy rule in the regime of high temperature. 

There are  many physical concepts  affecting human imagination ranging from confining light black holes, dilatation of time, butterfly effect up to teleportation and the celebrated but piteous Schr\"{o}dinger's cat. All of them are strange but the closed time--like curves are stranger than the other. We hope that our work will modestly contribute to both better understanding hypothetical behavior of quantum systems in a presence of CTCs and, as a guideline, to experimental attempts of mimicking such systems.  

\section*{Acknowledgments}
The  work has been  supported  by  the  NCN project UMO-2013/09/B/ST2/03382 (B.D. and M.R.) and the NCN grant 2015/19/B/ST2/02856 (J.D) 
%$ and the Forszt project co-financed by EU from the European Social Fund (M{\L}).

\section*{Appendix}
\begin{widetext}
In this Appendix we provide detail of calculations leading to the results presented in Sec III and V for the Deutchian model of CTC. Further in the Appendix we adopt the following notation
%%
%\begin{eqnarray}\label{not}
$p(i,j)=|i\rangle\langle j|, 
p(ij,kl)=|ij\rangle\langle kl|$ and $p(ijm,kln)=|ijm\rangle\langle kln|$
%\end{eqnarray}
%%
where $i,j,k,l,m,n=0,1$ labels our computational basis. In this notation  
partial traces of a two--qubit matrix $X$ with respect to CR and CV qubits and for 
%\begin{eqnarray}
$x(ij,kl)= \mbox{Tr}(X p(kl,ij))$
%\end{eqnarray}
 read as follows:
%\begin{widetext}
\begin{eqnarray}
\mbox{Tr}_{CR}X &=& (x(00,00)+x(10,10))p(0,0)+(x(01,00)+x(11,10))p(1,0)\nonumber \\&+&(x(00,01)+x(10,11))p(0,1)+(x(01,01)+x(11,11))p(1,1) \\
\mbox{Tr}_{CV}X &=& (x(00,00)+x(01,01))p(0,0)+(x(10,00)+x(11,01))p(1,0)\nonumber \\&+&(x(00,10)+x(01,11))p(0,1)+(x(10,10)+x(11,11))p(1,1)
\end{eqnarray}
%\end{widetext}
and the partial traces of a three--qubit matrix $X$ (in Sec. V) with respect to CR qubits and for 
%\begin{eqnarray}
$x(ijm,kln)= \mbox{Tr}(X p(klm,ijn))$
%\end{eqnarray}
 read as follows:
%\begin{widetext}
\begin{eqnarray}\label{tr3}
\mbox{Tr}_{CR}X &=& (x(000,000)+x(010,010)+x(100,100)+x(110,110))p(0,0)\nonumber\\&+&(x(000,001)+x(010,011)+x(100,101)+x(110,111))p(0,1)\nonumber\\&+&(x(001,000)+x(011,010)+x(101,100)+x(111,110))p(0,1)\nonumber\\&+&(x(001,001)+x(011,011)+x(101,101)+x(111,111))p(1,1)\end{eqnarray}

The unitary coupling between the CV and CR qubits  is a product of controlled Hadamard $H_C$
and the $SWAP$ i.e.
%\begin{widetext}
%\begin{eqnarray}
$U=H_C\,SWAP$ 
%\end{eqnarray}
where
%\begin{widetext}
\begin{eqnarray}
H_C &=& p(0,0)\otimes \mathcal{I} + p(1,1) \otimes Had\\
Had &=& (p(0,0)+p(0,1)+p(1,0)-p(1,1))/\sqrt{2}\\
SWAP &=& p(00,00)+p(11,11)+p(10,01)+p(01,10)
\end{eqnarray}
%\end{widetext}
For an input
$\rho_i=|-\rangle\langle -|=[p(0,0)+p(1,1)-p(1,0)-p(0,1)]/2$ the corresponding CV qubit 
%\begin{widetext}
\begin{eqnarray}\label{taug}
\tau&=&ap(0,0)+(1-a)p(1,1)+[(b_r+ib_i)p(0,1)+h.c.]
\end{eqnarray}
%\end{widetext}
 satisfying Eq.(\ref{d1}) with real $a,b_r,b_i$ can be calculated in the following steps:
{\it (i)} An output of the circuit $X=U \rho_i\otimes \tau U^\dagger$ is traced with respect to the CR qubit and then {\it (ii)} subjected to thermal noise via Eq.(\ref{dav}) and finally {\it (iii)} selfconsistently compared to the input i.e.:
%\begin{widetext}
\begin{eqnarray}
\tau &=& (x(00,00)+x(10,10))D[p(0,0)]+(x(01,00)+x(11,10))D[p(1,0)]\nonumber \\&+&(x(00,01)+x(10,11))D[p(0,1)]+(x(01,01)+x(11,11))D[p(1,1)]
\end{eqnarray}
%\end{widetext} 
resulting in a set of {\it linear} equations which allows to calculate the parameters $a,b_r,b_i$.
The CV qubit $\tau=\tau[1,1]p(0,0)+\tau[1,2]p(0,1)+\tau[2,1]p(1,0)+\tau[2,2]p(1,1)$ is then  given by
%
%\begin{widetext}
\begin{eqnarray}\label{dd2}
\tau[1,1]&=&1-\tau[2,2]=-2\,{\frac {p{{\rm e}^{-At}}-{{\rm e}^{-At}}-p+1}{{{\rm e}^{-At}}-2}}\nonumber \\
\tau[1,2]&=&\tau[2,1]^*={\frac {p{{\rm e}^{-t \left( i\omega+A+G \right) }}-{{\rm e}^{-t \left( i\omega+
G \right) }}p-{{\rm e}^{-t \left( i\omega+A+G \right) }}+{{\rm e}^{-t
 \left( i\omega+G \right) }}}{{{\rm e}^{-At}}-2}}
\end{eqnarray}
and the output $\rho_f$ of the circuit calculated via Eq.(\ref{d2}) reads as follows
\begin{eqnarray}\label{dd1}
\rho_f[1,1]&=&1-\rho_f[2,2]=-2\,{\frac {p{{\rm e}^{-At}}-{{\rm e}^{-At}}-p+1}{{{\rm e}^{-At}}-2}}\nonumber \\
\rho_f[1,2]&=&\rho_f[2,1]^*=
-1/2\,{\frac { \left( p{{\rm e}^{-t \left( i\omega+A+G \right) }}-{{\rm e}^
{-t \left( i\omega+G \right) }}p-{{\rm e}^{-t \left( i\omega+A+G \right) }}+{
{\rm e}^{-t \left( i\omega+G \right) }} \right) \sqrt {2}}{{{\rm e}^{-At}}-
2}}
\end{eqnarray}
For an input $\rho_i=|0\rangle\langle 0|=p(0,0)$
the CV qubit, calculated via the same steps, 
is given by:
%
%\begin{widetext}
%
%\begin{widetext}
\begin{eqnarray}\label{ddd2}
\tau[1,1]&=&1-\tau[2,2]=-{\frac {2\,p{{\rm e}^{-At}}-{{\rm e}^{-At}}-2\,p+2}{{{\rm e}^{-At}}-2
}}
\nonumber \\
\tau[1,2]&=&\tau[2,1]^*={\frac {p \left( {{\rm e}^{-t \left( i\omega+A+G \right) }}-{{\rm e}^{-t
 \left( i\omega+G \right) }} \right) }{{{\rm e}^{-At}}-2}}
\end{eqnarray}
and the corresponding output of the circuit calculated via Eq.(\ref{d2}) reads as follows
\begin{eqnarray}\label{ddd1}
\rho_f[1,1]&=&1-\rho_f[2,2]=-{\frac {2\,p{{\rm e}^{-At}}-{{\rm e}^{-At}}-2\,p+2}{{{\rm e}^{-At}}-2
}}
\nonumber \\
\rho_f[1,2]&=&\rho_f[2,1]^*=1/2\,{\frac {p \left( {{\rm e}^{-t \left( i\omega+A+G \right) }}-{{\rm e}^{
-t \left( i\omega+G \right) }} \right) \sqrt {2}}{{{\rm e}^{-At}}-2}}
\end{eqnarray}

In the case of the unproven theorem paradox considered in Sec. V the circuit
acts as a unitary 
\begin{eqnarray}
U&=& SWAP_{MT}CNOT_{BM}CNOT_{TB}
\end{eqnarray}
 with
\begin{eqnarray}
CNOT_{TB}&=&p(000,000)+p(101,001)+p(100,100)\nonumber \\ &+&
p(111,011)+p(001,101)+p(110,110)+
p(011,111)
\end{eqnarray}
\begin{eqnarray}
CNOT_{BM}&=&p(000,000)+
p(001,001)+p(010,010)+p(110,100)\nonumber\\&+&
p(011,011)+p(111,101)+p(100,110)+
p(101,111)
\end{eqnarray}
and
\begin{eqnarray}
SWAP_{MT}&=&p(000,000)+
p(010,001)+p(001,010)+p(100,100)\nonumber \\&+&
p(011,011)+p(110,101)+p(101,110)+
p(111,111)
\end{eqnarray}
 
An input $\rho_i=p(00,00)$ and $\tau$ is given in Eq.(\ref{taug}). Further one follows the same steps as in the previous case  but with $X=U \rho_i\otimes \tau U^\dagger$ then traced with respect to the CR qubits according to Eq.(\ref{tr3}) and subjected to thermal noise via Eq.(\ref{dav}) obtaining $\tau[1,1]=1-\tau[2,2]=p$ and $\tau[1,2]=\tau[2,1]=0$ corresponding to the Gibbs state of the time travelling qubit.
\end{widetext}

\bibliographystyle{apsrev4-1}
%\bibliography{LRref}



%\input{j2.bbl}
%\bibliography{mybib}

\begin{thebibliography}{35}%
\makeatletter
\providecommand \@ifxundefined [1]{%
 \@ifx{#1\undefined}
}%
\providecommand \@ifnum [1]{%
 \ifnum #1\expandafter \@firstoftwo
 \else \expandafter \@secondoftwo
 \fi
}%
\providecommand \@ifx [1]{%
 \ifx #1\expandafter \@firstoftwo
 \else \expandafter \@secondoftwo
 \fi
}%
\providecommand \natexlab [1]{#1}%
\providecommand \enquote  [1]{``#1''}%
\providecommand \bibnamefont  [1]{#1}%
\providecommand \bibfnamefont [1]{#1}%
\providecommand \citenamefont [1]{#1}%
\providecommand \href@noop [0]{\@secondoftwo}%
\providecommand \href [0]{\begingroup \@sanitize@url \@href}%
\providecommand \@href[1]{\@@startlink{#1}\@@href}%
\providecommand \@@href[1]{\endgroup#1\@@endlink}%
\providecommand \@sanitize@url [0]{\catcode `\\12\catcode `\$12\catcode
  `\&12\catcode `\#12\catcode `\^12\catcode `\_12\catcode `\%12\relax}%
\providecommand \@@startlink[1]{}%
\providecommand \@@endlink[0]{}%
\providecommand \url  [0]{\begingroup\@sanitize@url \@url }%
\providecommand \@url [1]{\endgroup\@href {#1}{\urlprefix }}%
\providecommand \urlprefix  [0]{URL }%
\providecommand \Eprint [0]{\href }%
\providecommand \doibase [0]{http://dx.doi.org/}%
\providecommand \selectlanguage [0]{\@gobble}%
\providecommand \bibinfo  [0]{\@secondoftwo}%
\providecommand \bibfield  [0]{\@secondoftwo}%
\providecommand \translation [1]{[#1]}%
\providecommand \BibitemOpen [0]{}%
\providecommand \bibitemStop [0]{}%
\providecommand \bibitemNoStop [0]{.\EOS\space}%
\providecommand \EOS [0]{\spacefactor3000\relax}%
\providecommand \BibitemShut  [1]{\csname bibitem#1\endcsname}%
\let\auto@bib@innerbib\@empty
%</preamble>
\bibitem [{\citenamefont {Nielsen}\ and\ \citenamefont
  {I.~Chuang}(2000)}]{nielsen}%
  \BibitemOpen
  \bibfield  {author} {\bibinfo {author} {\bibfnamefont {M.}~\bibnamefont
  {Nielsen}}\ and\ \bibinfo {author} {\bibfnamefont {I.}~\bibnamefont
  {I.~Chuang}},\ }\href@noop {} {\emph {\bibinfo {title} {Quantum Computation
  and Quantum Information}}}\ (\bibinfo  {publisher} {Cambridge University
  Press},\ \bibinfo {year} {2000})\BibitemShut {NoStop}%
\bibitem [{\citenamefont {Scarani}\ \emph {et~al.}(2009)\citenamefont
  {Scarani}, \citenamefont {Bechmann-Pasquinucci}, \citenamefont {Cerf},
  \citenamefont {Du\ifmmode~\check{s}\else \v{s}\fi{}ek}, \citenamefont
  {L\"utkenhaus},\ and\ \citenamefont {Peev}}]{scarani2}%
  \BibitemOpen
  \bibfield  {author} {\bibinfo {author} {\bibfnamefont {V.}~\bibnamefont
  {Scarani}}, \bibinfo {author} {\bibfnamefont {H.}~\bibnamefont
  {Bechmann-Pasquinucci}}, \bibinfo {author} {\bibfnamefont {N.~J.}\
  \bibnamefont {Cerf}}, \bibinfo {author} {\bibfnamefont {M.}~\bibnamefont
  {Du\ifmmode~\check{s}\else \v{s}\fi{}ek}}, \bibinfo {author} {\bibfnamefont
  {N.}~\bibnamefont {L\"utkenhaus}}, \ and\ \bibinfo {author} {\bibfnamefont
  {M.}~\bibnamefont {Peev}},\ }\href {\doibase 10.1103/RevModPhys.81.1301}
  {\bibfield  {journal} {\bibinfo  {journal} {Rev. Mod. Phys.}\ }\textbf
  {\bibinfo {volume} {81}},\ \bibinfo {pages} {1301} (\bibinfo {year}
  {2009})}\BibitemShut {NoStop}%
\bibitem [{\citenamefont {Gisin}\ \emph {et~al.}(2002)\citenamefont {Gisin},
  \citenamefont {Ribordy}, \citenamefont {Tittel},\ and\ \citenamefont
  {Zbinden}}]{crypto}%
  \BibitemOpen
  \bibfield  {author} {\bibinfo {author} {\bibfnamefont {N.}~\bibnamefont
  {Gisin}}, \bibinfo {author} {\bibfnamefont {G.}~\bibnamefont {Ribordy}},
  \bibinfo {author} {\bibfnamefont {W.}~\bibnamefont {Tittel}}, \ and\ \bibinfo
  {author} {\bibfnamefont {H.}~\bibnamefont {Zbinden}},\ }\href {\doibase
  10.1103/RevModPhys.74.145} {\bibfield  {journal} {\bibinfo  {journal} {Rev.
  Mod. Phys.}\ }\textbf {\bibinfo {volume} {74}},\ \bibinfo {pages} {145}
  (\bibinfo {year} {2002})}\BibitemShut {NoStop}%
\bibitem [{\citenamefont {G\"odel}(1949)}]{godel}%
  \BibitemOpen
  \bibfield  {author} {\bibinfo {author} {\bibfnamefont {K.}~\bibnamefont
  {G\"odel}},\ }\href {\doibase 10.1103/RevModPhys.21.447} {\bibfield
  {journal} {\bibinfo  {journal} {Rev. Mod. Phys.}\ }\textbf {\bibinfo {volume}
  {21}},\ \bibinfo {pages} {447} (\bibinfo {year} {1949})}\BibitemShut
  {NoStop}%
\bibitem [{\citenamefont {Brun}\ and\ \citenamefont {Wilde}(2015)}]{brun_exp}%
  \BibitemOpen
  \bibfield  {author} {\bibinfo {author} {\bibfnamefont {T.~A.}\ \bibnamefont
  {Brun}}\ and\ \bibinfo {author} {\bibfnamefont {M.~M.}\ \bibnamefont
  {Wilde}},\ }\href {http://arxiv.org/abs/1504.05911} {\bibfield  {journal}
  {\bibinfo  {journal} {arxiv.org/abs/1504.05911}\ } (\bibinfo {year}
  {2015})}\BibitemShut {NoStop}%
\bibitem [{\citenamefont {Brun}\ \emph {et~al.}(2009)\citenamefont {Brun},
  \citenamefont {Harrington},\ and\ \citenamefont {Wilde}}]{brun_disting}%
  \BibitemOpen
  \bibfield  {author} {\bibinfo {author} {\bibfnamefont {T.~A.}\ \bibnamefont
  {Brun}}, \bibinfo {author} {\bibfnamefont {J.}~\bibnamefont {Harrington}}, \
  and\ \bibinfo {author} {\bibfnamefont {M.~M.}\ \bibnamefont {Wilde}},\ }\href
  {\doibase 10.1103/PhysRevLett.102.210402} {\bibfield  {journal} {\bibinfo
  {journal} {Phys. Rev. Lett.}\ }\textbf {\bibinfo {volume} {102}},\ \bibinfo
  {pages} {210402} (\bibinfo {year} {2009})}\BibitemShut {NoStop}%
\bibitem [{\citenamefont {Brun}\ and\ \citenamefont {Wilde}(2012)}]{brun_fund}%
  \BibitemOpen
  \bibfield  {author} {\bibinfo {author} {\bibfnamefont {T.~A.}\ \bibnamefont
  {Brun}}\ and\ \bibinfo {author} {\bibfnamefont {M.~M.}\ \bibnamefont
  {Wilde}},\ }\href@noop {} {\bibfield  {journal} {\bibinfo  {journal} {Found.
  Phys.}\ }\textbf {\bibinfo {volume} {42}},\ \bibinfo {pages} {341} (\bibinfo
  {year} {2012})}\BibitemShut {NoStop}%
\bibitem [{\citenamefont {Deutsch}(1991)}]{deutsch}%
  \BibitemOpen
  \bibfield  {author} {\bibinfo {author} {\bibfnamefont {D.}~\bibnamefont
  {Deutsch}},\ }\href {\doibase 10.1103/PhysRevD.44.3197} {\bibfield  {journal}
  {\bibinfo  {journal} {Phys. Rev. D}\ }\textbf {\bibinfo {volume} {44}},\
  \bibinfo {pages} {3197} (\bibinfo {year} {1991})}\BibitemShut {NoStop}%
\bibitem [{\citenamefont {Ringbauer}\ \emph {et~al.}(2014)\citenamefont
  {Ringbauer}, \citenamefont {Broome}, \citenamefont {Myers}, \citenamefont
  {White},\ and\ \citenamefont {Ralph}}]{ralph}%
  \BibitemOpen
  \bibfield  {author} {\bibinfo {author} {\bibfnamefont {M.}~\bibnamefont
  {Ringbauer}}, \bibinfo {author} {\bibfnamefont {M.~A.}\ \bibnamefont
  {Broome}}, \bibinfo {author} {\bibfnamefont {C.~R.}\ \bibnamefont {Myers}},
  \bibinfo {author} {\bibfnamefont {A.~G.}\ \bibnamefont {White}}, \ and\
  \bibinfo {author} {\bibfnamefont {T.~C.}\ \bibnamefont {Ralph}},\ }\href@noop
  {} {\bibfield  {journal} {\bibinfo  {journal} {Nature Communications}\
  }\textbf {\bibinfo {volume} {5}},\ \bibinfo {pages} {4145} (\bibinfo {year}
  {2014})}\BibitemShut {NoStop}%
\bibitem [{\citenamefont {Wallman}\ and\ \citenamefont
  {Bartlett}(2012)}]{wal_fund}%
  \BibitemOpen
  \bibfield  {author} {\bibinfo {author} {\bibfnamefont {J.~J.}\ \bibnamefont
  {Wallman}}\ and\ \bibinfo {author} {\bibfnamefont {S.~D.}\ \bibnamefont
  {Bartlett}},\ }\href@noop {} {\bibfield  {journal} {\bibinfo  {journal}
  {Found. Phys.}\ }\textbf {\bibinfo {volume} {42}},\ \bibinfo {pages} {656}
  (\bibinfo {year} {2012})}\BibitemShut {NoStop}%
\bibitem [{\citenamefont {Allen}(2014)}]{allen}%
  \BibitemOpen
  \bibfield  {author} {\bibinfo {author} {\bibfnamefont {J.-M.~A.}\
  \bibnamefont {Allen}},\ }\href {\doibase 10.1103/PhysRevA.90.042107}
  {\bibfield  {journal} {\bibinfo  {journal} {Phys. Rev. A}\ }\textbf {\bibinfo
  {volume} {90}},\ \bibinfo {pages} {042107} (\bibinfo {year}
  {2014})}\BibitemShut {NoStop}%
\bibitem [{\citenamefont {Svetlichny}(2011)}]{svet}%
  \BibitemOpen
  \bibfield  {author} {\bibinfo {author} {\bibfnamefont {G.}~\bibnamefont
  {Svetlichny}},\ }\href {\doibase 10.1007/s10773-011-0973-x} {\bibfield
  {journal} {\bibinfo  {journal} {International Journal of Theoretical
  Physics}\ }\textbf {\bibinfo {volume} {50}},\ \bibinfo {pages} {3903}
  (\bibinfo {year} {2011})}\BibitemShut {NoStop}%
\bibitem [{\citenamefont {Lloyd}\ \emph
  {et~al.}(2011{\natexlab{a}})\citenamefont {Lloyd}, \citenamefont {Maccone},
  \citenamefont {Garcia-Patron}, \citenamefont {Giovannetti}, \citenamefont
  {Shikano}, \citenamefont {Pirandola}, \citenamefont {Rozema}, \citenamefont
  {Darabi}, \citenamefont {Soudagar}, \citenamefont {Shalm},\ and\
  \citenamefont {Steinberg}}]{seth_prl}%
  \BibitemOpen
  \bibfield  {author} {\bibinfo {author} {\bibfnamefont {S.}~\bibnamefont
  {Lloyd}}, \bibinfo {author} {\bibfnamefont {L.}~\bibnamefont {Maccone}},
  \bibinfo {author} {\bibfnamefont {R.}~\bibnamefont {Garcia-Patron}}, \bibinfo
  {author} {\bibfnamefont {V.}~\bibnamefont {Giovannetti}}, \bibinfo {author}
  {\bibfnamefont {Y.}~\bibnamefont {Shikano}}, \bibinfo {author} {\bibfnamefont
  {S.}~\bibnamefont {Pirandola}}, \bibinfo {author} {\bibfnamefont {L.~A.}\
  \bibnamefont {Rozema}}, \bibinfo {author} {\bibfnamefont {A.}~\bibnamefont
  {Darabi}}, \bibinfo {author} {\bibfnamefont {Y.}~\bibnamefont {Soudagar}},
  \bibinfo {author} {\bibfnamefont {L.~K.}\ \bibnamefont {Shalm}}, \ and\
  \bibinfo {author} {\bibfnamefont {A.~M.}\ \bibnamefont {Steinberg}},\ }\href
  {\doibase 10.1103/PhysRevLett.106.040403} {\bibfield  {journal} {\bibinfo
  {journal} {Phys. Rev. Lett.}\ }\textbf {\bibinfo {volume} {106}},\ \bibinfo
  {pages} {040403} (\bibinfo {year} {2011}{\natexlab{a}})}\BibitemShut
  {NoStop}%
\bibitem [{\citenamefont {Lloyd}\ \emph
  {et~al.}(2011{\natexlab{b}})\citenamefont {Lloyd}, \citenamefont {Maccone},
  \citenamefont {Garcia-Patron}, \citenamefont {Giovannetti},\ and\
  \citenamefont {Shikano}}]{seth_prd}%
  \BibitemOpen
  \bibfield  {author} {\bibinfo {author} {\bibfnamefont {S.}~\bibnamefont
  {Lloyd}}, \bibinfo {author} {\bibfnamefont {L.}~\bibnamefont {Maccone}},
  \bibinfo {author} {\bibfnamefont {R.}~\bibnamefont {Garcia-Patron}}, \bibinfo
  {author} {\bibfnamefont {V.}~\bibnamefont {Giovannetti}}, \ and\ \bibinfo
  {author} {\bibfnamefont {Y.}~\bibnamefont {Shikano}},\ }\href {\doibase
  10.1103/PhysRevD.84.025007} {\bibfield  {journal} {\bibinfo  {journal} {Phys.
  Rev. D}\ }\textbf {\bibinfo {volume} {84}},\ \bibinfo {pages} {025007}
  (\bibinfo {year} {2011}{\natexlab{b}})}\BibitemShut {NoStop}%
\bibitem [{\citenamefont {{Elze, Hans-Thomas}}(2013)}]{elze_time}%
  \BibitemOpen
  \bibfield  {author} {\bibinfo {author} {\bibnamefont {{Elze, Hans-Thomas}}},\
  }\href {\doibase 10.1051/epjconf/20135801013} {\bibfield  {journal} {\bibinfo
   {journal} {EPJ Web of Conferences}\ }\textbf {\bibinfo {volume} {58}},\
  \bibinfo {pages} {01013} (\bibinfo {year} {2013})}\BibitemShut {NoStop}%
\bibitem [{\citenamefont {Vaidman}(2013)}]{vaidman_past}%
  \BibitemOpen
  \bibfield  {author} {\bibinfo {author} {\bibfnamefont {L.}~\bibnamefont
  {Vaidman}},\ }\href {\doibase 10.1103/PhysRevA.87.052104} {\bibfield
  {journal} {\bibinfo  {journal} {Phys. Rev. A}\ }\textbf {\bibinfo {volume}
  {87}},\ \bibinfo {pages} {052104} (\bibinfo {year} {2013})}\BibitemShut
  {NoStop}%
\bibitem [{\citenamefont {Alicki}\ and\ \citenamefont {Lendi}(2007)}]{alicki}%
  \BibitemOpen
  \bibfield  {author} {\bibinfo {author} {\bibfnamefont {R.}~\bibnamefont
  {Alicki}}\ and\ \bibinfo {author} {\bibfnamefont {K.}~\bibnamefont {Lendi}},\
  }\href@noop {} {\emph {\bibinfo {title} {Quantum Dynamical Semigroups and
  Applications}}},\ Lecture Notes in Physics\ (\bibinfo  {publisher}
  {Springer},\ \bibinfo {year} {2007})\BibitemShut {NoStop}%
\bibitem [{\citenamefont {K{\l}oda}\ and\ \citenamefont {Dajka}(2014)}]{kloda}%
  \BibitemOpen
  \bibfield  {author} {\bibinfo {author} {\bibfnamefont {D.}~\bibnamefont
  {K{\l}oda}}\ and\ \bibinfo {author} {\bibfnamefont {J.}~\bibnamefont
  {Dajka}},\ }\href {\doibase 10.1007/s11128-014-0831-x} {\bibfield  {journal}
  {\bibinfo  {journal} {Quantum Information Processing}\ ,\ \bibinfo {pages}
  {1}} (\bibinfo {year} {2014})}\BibitemShut {NoStop}%
\bibitem [{\citenamefont {Lendi}\ and\ \citenamefont {Wonderen}(2007)}]{lendi}%
  \BibitemOpen
  \bibfield  {author} {\bibinfo {author} {\bibfnamefont {K.}~\bibnamefont
  {Lendi}}\ and\ \bibinfo {author} {\bibfnamefont {A.~J.~v.}\ \bibnamefont
  {Wonderen}},\ }\href@noop {} {\bibfield  {journal} {\bibinfo  {journal}
  {Journal of Physics A: Mathematical and Theoretical}\ }\textbf {\bibinfo
  {volume} {40}},\ \bibinfo {pages} {279} (\bibinfo {year} {2007})}\BibitemShut
  {NoStop}%
\bibitem [{\citenamefont {Dajka}\ \emph {et~al.}(2012)\citenamefont {Dajka},
  \citenamefont {Mierzejewski}, \citenamefont {\L{}uczka}, \citenamefont
  {Blattmann},\ and\ \citenamefont {H{\"a}nggi}}]{mymy}%
  \BibitemOpen
  \bibfield  {author} {\bibinfo {author} {\bibfnamefont {J.}~\bibnamefont
  {Dajka}}, \bibinfo {author} {\bibfnamefont {M.}~\bibnamefont {Mierzejewski}},
  \bibinfo {author} {\bibfnamefont {J.}~\bibnamefont {\L{}uczka}}, \bibinfo
  {author} {\bibfnamefont {R.}~\bibnamefont {Blattmann}}, \ and\ \bibinfo
  {author} {\bibfnamefont {P.}~\bibnamefont {H{\"a}nggi}},\ }\href@noop {}
  {\bibfield  {journal} {\bibinfo  {journal} {Journal of Physics A:
  Mathematical and Theoretical}\ }\textbf {\bibinfo {volume} {45}},\ \bibinfo
  {pages} {485306} (\bibinfo {year} {2012})}\BibitemShut {NoStop}%
\bibitem [{\citenamefont {Dajka}\ and\ \citenamefont
  {\L{}uczka}(2013)}]{mymy2}%
  \BibitemOpen
  \bibfield  {author} {\bibinfo {author} {\bibfnamefont {J.}~\bibnamefont
  {Dajka}}\ and\ \bibinfo {author} {\bibfnamefont {J.}~\bibnamefont
  {\L{}uczka}},\ }\href@noop {} {\bibfield  {journal} {\bibinfo  {journal}
  {Phys. Rev. A}\ }\textbf {\bibinfo {volume} {87}},\ \bibinfo {pages} {022301}
  (\bibinfo {year} {2013})}\BibitemShut {NoStop}%
\bibitem [{\citenamefont {Dajka}\ \emph {et~al.}(2011)\citenamefont {Dajka},
  \citenamefont {\L{}uczka},\ and\ \citenamefont {H{\"a}nggi}}]{dav_faza}%
  \BibitemOpen
  \bibfield  {author} {\bibinfo {author} {\bibfnamefont {J.}~\bibnamefont
  {Dajka}}, \bibinfo {author} {\bibfnamefont {J.}~\bibnamefont {\L{}uczka}}, \
  and\ \bibinfo {author} {\bibfnamefont {P.}~\bibnamefont {H{\"a}nggi}},\
  }\href@noop {} {\bibfield  {journal} {\bibinfo  {journal} {Quantum
  Information Processing}\ }\textbf {\bibinfo {volume} {10}},\ \bibinfo {pages}
  {85} (\bibinfo {year} {2011})}\BibitemShut {NoStop}%
\bibitem [{\citenamefont {Szel{\c a}g}\ \emph {et~al.}(2008)\citenamefont
  {Szel{\c a}g}, \citenamefont {Dajka}, \citenamefont {Zipper},\ and\
  \citenamefont {\L{}uczka}}]{dav_heat}%
  \BibitemOpen
  \bibfield  {author} {\bibinfo {author} {\bibfnamefont {M.}~\bibnamefont
  {Szel{\c a}g}}, \bibinfo {author} {\bibfnamefont {J.}~\bibnamefont {Dajka}},
  \bibinfo {author} {\bibfnamefont {E.}~\bibnamefont {Zipper}}, \ and\ \bibinfo
  {author} {\bibfnamefont {J.}~\bibnamefont {\L{}uczka}},\ }\href@noop {}
  {\bibfield  {journal} {\bibinfo  {journal} {Acta Physica Polonica B}\
  }\textbf {\bibinfo {volume} {39}},\ \bibinfo {pages} {1177} (\bibinfo {year}
  {2008})}\BibitemShut {NoStop}%
\bibitem [{\citenamefont {Dajka}\ \emph {et~al.}(2015)\citenamefont {Dajka},
  \citenamefont {K{\l}oda}, \citenamefont {{\L}obejko},\ and\ \citenamefont
  {S{\l}adkowski}}]{dajka_game}%
  \BibitemOpen
  \bibfield  {author} {\bibinfo {author} {\bibfnamefont {J.}~\bibnamefont
  {Dajka}}, \bibinfo {author} {\bibfnamefont {D.}~\bibnamefont {K{\l}oda}},
  \bibinfo {author} {\bibfnamefont {M.}~\bibnamefont {{\L}obejko}}, \ and\
  \bibinfo {author} {\bibfnamefont {J.}~\bibnamefont {S{\l}adkowski}},\ }\href
  {\doibase 10.1371/journal.pone.0134916} {\bibfield  {journal} {\bibinfo
  {journal} {PLoS ONE}\ }\textbf {\bibinfo {volume} {10}},\ \bibinfo {pages}
  {1} (\bibinfo {year} {2015})}\BibitemShut {NoStop}%
\bibitem [{\citenamefont {Roga}\ \emph {et~al.}(2010)\citenamefont {Roga},
  \citenamefont {Fannes},\ and\ \citenamefont {Zyczkowski}}]{dav}%
  \BibitemOpen
  \bibfield  {author} {\bibinfo {author} {\bibfnamefont {W.}~\bibnamefont
  {Roga}}, \bibinfo {author} {\bibfnamefont {M.}~\bibnamefont {Fannes}}, \ and\
  \bibinfo {author} {\bibfnamefont {K.}~\bibnamefont {Zyczkowski}},\
  }\href@noop {} {\bibfield  {journal} {\bibinfo  {journal} {Reports on
  Mathematical Physics}\ }\textbf {\bibinfo {volume} {66}},\ \bibinfo {pages}
  {311 } (\bibinfo {year} {2010})}\BibitemShut {NoStop}%
\bibitem [{\citenamefont {Levitt}(2008)}]{T12}%
  \BibitemOpen
  \bibfield  {author} {\bibinfo {author} {\bibfnamefont {M.~H.}\ \bibnamefont
  {Levitt}},\ }\href@noop {} {\emph {\bibinfo {title} {Spin Dynamics: Basics of
  Nuclear Magnetic Resonance}}}\ (\bibinfo  {publisher} {Wiley},\ \bibinfo
  {year} {2008})\BibitemShut {NoStop}%
\bibitem [{\citenamefont {Schuster}\ \emph {et~al.}(2007)\citenamefont
  {Schuster}, \citenamefont {Houck}, \citenamefont {Schreier}, \citenamefont
  {Wallraff}, \citenamefont {Gambetta}, \citenamefont {Blais}, \citenamefont
  {Frunzio}, \citenamefont {Majer}, \citenamefont {Johnson}, \citenamefont
  {Devoret}, \citenamefont {Girvin},\ and\ \citenamefont {Schoelkopf}}]{defaz}%
  \BibitemOpen
  \bibfield  {author} {\bibinfo {author} {\bibfnamefont {D.~I.}\ \bibnamefont
  {Schuster}}, \bibinfo {author} {\bibfnamefont {A.~A.}\ \bibnamefont {Houck}},
  \bibinfo {author} {\bibfnamefont {J.~A.}\ \bibnamefont {Schreier}}, \bibinfo
  {author} {\bibfnamefont {A.}~\bibnamefont {Wallraff}}, \bibinfo {author}
  {\bibfnamefont {J.~M.}\ \bibnamefont {Gambetta}}, \bibinfo {author}
  {\bibfnamefont {A.}~\bibnamefont {Blais}}, \bibinfo {author} {\bibfnamefont
  {L.}~\bibnamefont {Frunzio}}, \bibinfo {author} {\bibfnamefont
  {J.}~\bibnamefont {Majer}}, \bibinfo {author} {\bibfnamefont
  {B.}~\bibnamefont {Johnson}}, \bibinfo {author} {\bibfnamefont {M.~H.}\
  \bibnamefont {Devoret}}, \bibinfo {author} {\bibfnamefont {S.~M.}\
  \bibnamefont {Girvin}}, \ and\ \bibinfo {author} {\bibfnamefont {R.~J.}\
  \bibnamefont {Schoelkopf}},\ }\href@noop {} {\bibfield  {journal} {\bibinfo
  {journal} {Nature}\ }\textbf {\bibinfo {volume} {445}},\ \bibinfo {pages}
  {515} (\bibinfo {year} {2007})}\BibitemShut {NoStop}%
\bibitem [{\citenamefont {Bennett}(1992)}]{B92}%
  \BibitemOpen
  \bibfield  {author} {\bibinfo {author} {\bibfnamefont {C.~H.}\ \bibnamefont
  {Bennett}},\ }\href {\doibase 10.1103/PhysRevLett.68.3121} {\bibfield
  {journal} {\bibinfo  {journal} {Phys. Rev. Lett.}\ }\textbf {\bibinfo
  {volume} {68}},\ \bibinfo {pages} {3121} (\bibinfo {year}
  {1992})}\BibitemShut {NoStop}%
\bibitem [{\citenamefont {Alicki}(2004)}]{alidef}%
  \BibitemOpen
  \bibfield  {author} {\bibinfo {author} {\bibfnamefont {R.}~\bibnamefont
  {Alicki}},\ }\href@noop {} {\bibfield  {journal} {\bibinfo  {journal} {Open
  Syst. Inf. Dyn.}\ }\textbf {\bibinfo {volume} {11}},\ \bibinfo {pages} {53}
  (\bibinfo {year} {2004})}\BibitemShut {NoStop}%
\bibitem [{\citenamefont {Dajka}\ and\ \citenamefont
  {\L{}uczka}(2010)}]{distance}%
  \BibitemOpen
  \bibfield  {author} {\bibinfo {author} {\bibfnamefont {J.}~\bibnamefont
  {Dajka}}\ and\ \bibinfo {author} {\bibfnamefont {J.}~\bibnamefont
  {\L{}uczka}},\ }\href {\doibase 10.1103/PhysRevA.82.012341} {\bibfield
  {journal} {\bibinfo  {journal} {Phys. Rev. A}\ }\textbf {\bibinfo {volume}
  {82}},\ \bibinfo {pages} {012341} (\bibinfo {year} {2010})}\BibitemShut
  {NoStop}%
\bibitem [{\citenamefont {Laine}\ \emph {et~al.}(2010)\citenamefont {Laine},
  \citenamefont {Piilo},\ and\ \citenamefont {Breuer}}]{laine}%
  \BibitemOpen
  \bibfield  {author} {\bibinfo {author} {\bibfnamefont {E.-M.}\ \bibnamefont
  {Laine}}, \bibinfo {author} {\bibfnamefont {J.}~\bibnamefont {Piilo}}, \ and\
  \bibinfo {author} {\bibfnamefont {H.-P.}\ \bibnamefont {Breuer}},\ }\href
  {http://stacks.iop.org/0295-5075/92/i=6/a=60010} {\bibfield  {journal}
  {\bibinfo  {journal} {EPL (Europhysics Letters)}\ }\textbf {\bibinfo {volume}
  {92}},\ \bibinfo {pages} {60010} (\bibinfo {year} {2010})}\BibitemShut
  {NoStop}%
\bibitem [{\citenamefont {Politzer}(1994)}]{politzer}%
  \BibitemOpen
  \bibfield  {author} {\bibinfo {author} {\bibfnamefont {H.~D.}\ \bibnamefont
  {Politzer}},\ }\href {\doibase 10.1103/PhysRevD.49.3981} {\bibfield
  {journal} {\bibinfo  {journal} {Phys. Rev. D}\ }\textbf {\bibinfo {volume}
  {49}},\ \bibinfo {pages} {3981} (\bibinfo {year} {1994})}\BibitemShut
  {NoStop}%
\bibitem [{\citenamefont {DeJonghe}\ \emph {et~al.}(2010)\citenamefont
  {DeJonghe}, \citenamefont {Frey},\ and\ \citenamefont {Imbo}}]{dejonghe}%
  \BibitemOpen
  \bibfield  {author} {\bibinfo {author} {\bibfnamefont {R.}~\bibnamefont
  {DeJonghe}}, \bibinfo {author} {\bibfnamefont {K.}~\bibnamefont {Frey}}, \
  and\ \bibinfo {author} {\bibfnamefont {T.}~\bibnamefont {Imbo}},\ }\href
  {\doibase 10.1103/PhysRevD.81.087501} {\bibfield  {journal} {\bibinfo
  {journal} {Phys. Rev. D}\ }\textbf {\bibinfo {volume} {81}},\ \bibinfo
  {pages} {087501} (\bibinfo {year} {2010})}\BibitemShut {NoStop}%
\bibitem [{\citenamefont {Scarani}\ \emph {et~al.}(2005)\citenamefont
  {Scarani}, \citenamefont {Iblisdir}, \citenamefont {Gisin},\ and\
  \citenamefont {Ac\'{i}n}}]{scarani1}%
  \BibitemOpen
  \bibfield  {author} {\bibinfo {author} {\bibfnamefont {V.}~\bibnamefont
  {Scarani}}, \bibinfo {author} {\bibfnamefont {S.}~\bibnamefont {Iblisdir}},
  \bibinfo {author} {\bibfnamefont {N.}~\bibnamefont {Gisin}}, \ and\ \bibinfo
  {author} {\bibfnamefont {A.}~\bibnamefont {Ac\'{i}n}},\ }\href {\doibase
  10.1103/RevModPhys.77.1225} {\bibfield  {journal} {\bibinfo  {journal} {Rev.
  Mod. Phys.}\ }\textbf {\bibinfo {volume} {77}},\ \bibinfo {pages} {1225}
  (\bibinfo {year} {2005})}\BibitemShut {NoStop}%
\bibitem [{\citenamefont {Schlosshauer}(2007)}]{schloss}%
  \BibitemOpen
  \bibfield  {author} {\bibinfo {author} {\bibfnamefont {M.}~\bibnamefont
  {Schlosshauer}},\ }\href@noop {} {\emph {\bibinfo {title} {Decoherence and
  the quantum-to-classical transition}}}\ (\bibinfo  {publisher} {Springer},\
  \bibinfo {year} {2007})\BibitemShut {NoStop}%
\end{thebibliography}

\end{document}